\documentclass[twocolumn,a4paper,10pt]{article}
\usepackage{amsmath,amsfonts}
\usepackage{epsfig}
\usepackage{times}
\usepackage{epsfig}
\usepackage{booktabs}
\usepackage{graphicx}
\usepackage{url}
\usepackage{color}
\usepackage{tikz}
\usepackage[small,hang]{caption2}
\usepackage{color}
\makeatletter

\newcounter{numbersec}
\renewcommand{\section}[1]{\par\noindent\stepcounter{numbersec}
\par
\vspace{6pt}
\noindent\textbf{\large   \arabic{numbersec} \hspace*{0.3cm} #1 }
\par
\vspace{2pt}
}
\renewcommand{\subsection}[1]{
\par
\vspace{6pt}
\noindent\textbf{#1}
\par
}
\renewcommand{\subsubsection}[1]{%
\par
\vspace{6pt}
\textbf{#1.}
}

\newcommand{\Abstract}{\par\vspace{6pt}\noindent\textbf{\large Abstract}\par\vspace{2pt}}
\newcommand{\Acknowledgments}{\par\vspace{6pt}\noindent\textbf{\large Acknowledgments }\par\vspace{2pt}}

\newenvironment{References}{
\par\vspace{6pt}\noindent\textbf{\large References}\par\vspace{2pt}
\begin{small}\begin{list}{ }{
\itemsep0mm \parsep0mm\labelsep0mm\leftmargin0mm
}}
{\end{list}\end{small}}

\makeatother

\title{\vspace*{-12mm}
\LARGE \sc \textbf{  
Assessment of skin-friction-reduction techniques on a turbulent wing section
}}
\author{ \Large \bf \textit{M. Atzori$^{1*}$,} \textit{R. Vinuesa$^{1}$,} \textit{A. Stroh$^{2}$,} \textit{B. Frohnapfel$^{2}$ and} \textit{P. Schlatter$^{1}$}  \\ \\
\bf$^{1}$\textit{Linn\'e FLOW Centre, KTH Mechanics} \\
\bf \textit{and Swedish e-Science Research Centre (SeRC), SE-100 44, Stockholm, Sweden} \\ \\
\bf$^{2}$\textit{Institute of Fluid Mechanics, Karlsruhe Institute of Technology (KIT), Karlsruhe, Germany} \\ \\ 
*\underline{\bf atzori@mech.kth.se}
}
\date{}

\begin{document}
%


%

\maketitle
\thispagestyle{empty}



\newcommand{\tocheck}[1] {{\color{red} \textbf{#1}}}

%
%
\Abstract
The scope of the present project is to quantify the effects of uniform blowing and body-force damping on turbulent boundary layers subjected to a non-uniform adverse-pressure-gradient distribution. To this end, well-resolved large-eddy simulations are employed to describe the flow around the NACA4412 airfoil at moderate Reynolds number $200,000$ based on free-stream velocity and chord length. In the present paper we focus on uniform blowing and the conference presentation will include a comparison with body-force damping applied in the same region. The inner-scaled profiles of the mean velocity and of selected components of the Reynolds-stress tensor are examined and compared with the uncontrolled cases. It is known that uniform blowing and adverse-pressure gradients share some similarities in their effect on the boundary layers, and our results will show that these effects are not independent. The behaviour of the skin-friction coefficient is analyzed through the FIK decomposition, and the impact of this control strategy on the aerodynamic efficiency of the airfoil is discussed. 
%
%
\section{Introduction}
The development of effective techniques for the reduction of skin friction on airfoils remains a challenging task with relevant implications both from the economical and the environmental points of view. Estimates of the potential savings in industrial applications achievable by reducing the skin-friction drag have been given, for instance, by Gad-el-Hak (2000). Several strategies have been proposed in the last decades, including either passive (\emph{e.g.} riblets, superhydrophobic surfaces) or active (\emph{e.g.} blowing, wall oscillation) techniques, which have been studied both via experiments and numerical simulations (see the review of flow-control methods by Choi \emph{et al.}, 2008). However, simple test cases, such as turbulent channel flow or - more seldomly - turbulent boundary layers, have been employed in those numerical studies because of the high computational cost of these simulations. These results cannot be easily generalised to more relevant cases such as the flow around wing sections because of the combined effects of wall curvature and the streamwise pressure gradient. Moreover, to fully assess the effectiveness of a control strategy for practical aeronautical applications, it is necessary to take into account its overall impact on the aerodynamic properties of the wing under study. In the present paper we describe the effect of uniform blowing on the turbulent boundary layer over the suction side of a NACA4412 profile at a Reynolds number $Re_c=200,000$ (based on the inflow velocity $U_\infty$ and the chord length $c$). These results an extension of those of the preliminary study at $Re_c=100,000$ by Vinuesa and Schlatter (2017). 
\section{Numerical method}
\begin{figure}[htp]
\begin{center}
 \includegraphics[width=0.45\textwidth]{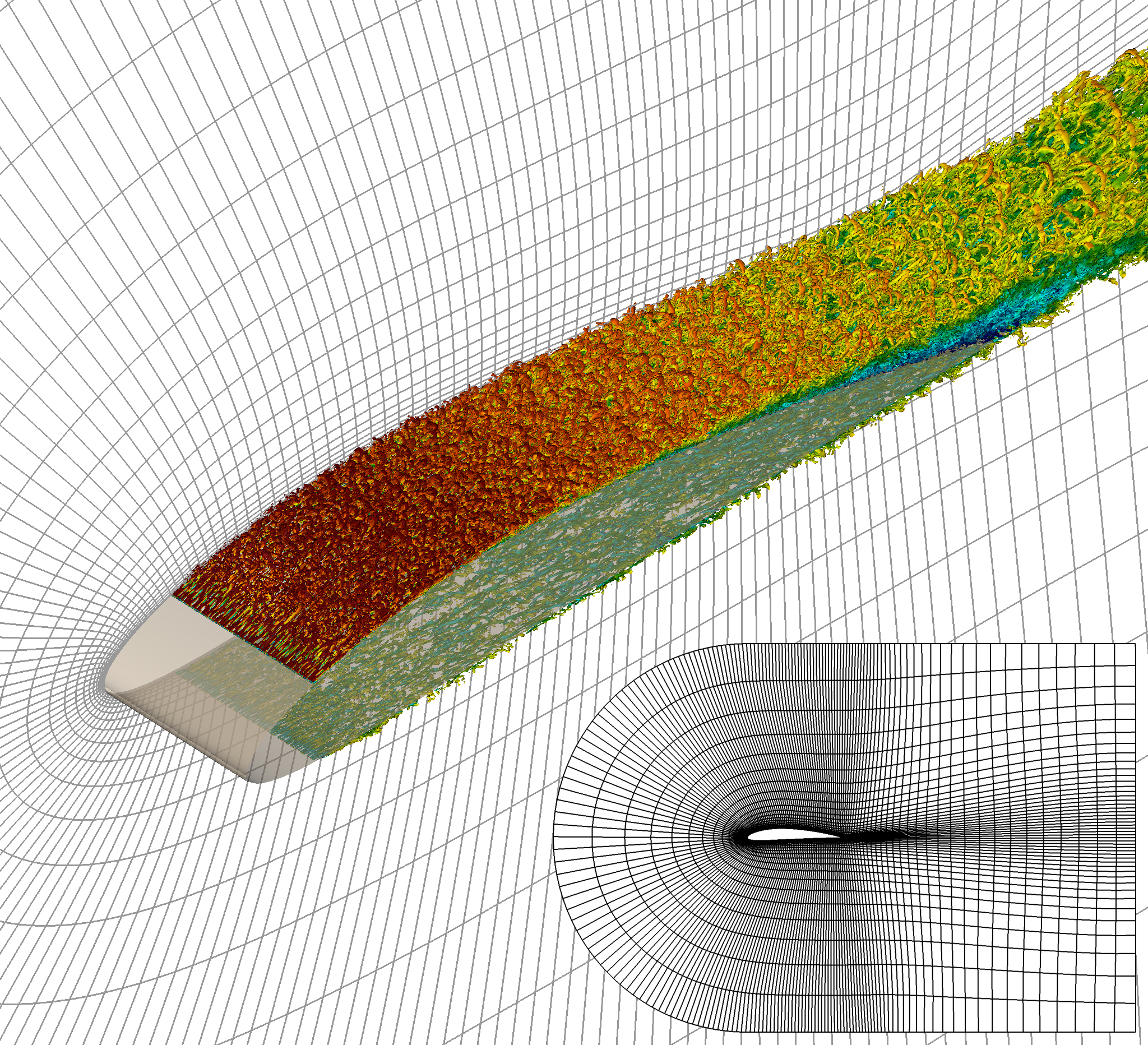}
\end{center}
 \caption{Spectral-element mesh and visualisation of vortex clusters for the $Re_c=200,000$ simulation. The vortex clusters are defined as an isosurface of $\lambda_2$ (Jeong and Hussain, 1995) and they are colored by their streamwise velocity, where dark red is $1.6 U_\infty$ and dark blue $-0.18 U_\infty$. The computational domain is shown in the insert.}
 \label{fig:wing}
\end{figure}

The set-up of the numerical simulations is analogous to the one described in details by Vinuesa \emph{et al.} (2018), who also present the results for the uncontrolled reference cases employed here, whereas the strategies for mesh design and computation of turbulence statistics are discussed by Vinuesa \emph{et al.} (2017). The code is \emph{Nek5000}, developed by Fischer \emph{et al.} (2008), which uses a spectral-element method (SEM). We adopted a C-mesh with total lengths of $L_{x}=6c$ and $L_{y}=4c$ in the streamwise and vertical directions, as can be seen in Fig.~\ref{fig:wing}. 
The width of the domain in the periodic spanwise direction is $L_z=0.2c$. The boundary conditions are imposed considering a RANS simulation on a domain extending to a length of $200c$ in the streamwise and wall-normal directions. The flow is tripped on both the sides of the airfoil, at a distance of $x/c=0.1$ from the leading edge. The tripping is implemented as a volumetric force in the vertical direction, which is designed to have the same effect as that of the tripping strips in experimental facilities (Schlatter and \"Orl\"u, 2012). The number of spectral elements is approximately $125,000$, which leads to $216$ million grid points for the considered polynomial order $N=11$. The resulting resolution in the tangential, wall-normal and spanwise directions is $\Delta x_t^+=18$, $\Delta y_n^+=(0.64;11)$, and $\Delta z^+=9$, respectively. The spacing is scaled in inner units, employing the viscous length $l^*=\nu/u_\tau$, where $\nu$ is the kinematic viscosity of the fluid, $u_\tau=\sqrt{\tau_w/\rho}$ is the friction velocity, $\tau_w=\rho \nu ({\rm d} U_t/ {\rm d} y_n)_{y_n=0}$ the wall-shear stress and $\rho$ the fluid density. We perform a well-resolved large-eddy simulation (LES) using an approach based an explicit regularisation using a relaxation term (ADM-RT), introduced by Schlatter \emph{et al.} (2004). The spectral-element mesh is shown in the insert in Fig.~\ref{fig:wing}, together with a rendering of coherent vortices. Uniform blowing is imposed through the change of the wall boundary condition, where a non-zero wall-normal velocity is imposed. The Reynolds number $Re_c=200,000$ has been chosen because it is considered an acceptable compromise between computational cost and boundary-layer features comparable with those at higher $Re_c$. With the present setup, approximately $10$ flow-over times are required to obtain statistical convergence for the physical quantities of interest, which requires approximately $1.5$ million CPU-hours on the HPC system \emph{Beskow} of the Swedish National Infrastructure for Computing at PDC, KTH. 
\section{Results and discussion}
\begin{figure}[htp]
\begin{center}
 \includegraphics[width=0.4\textwidth]{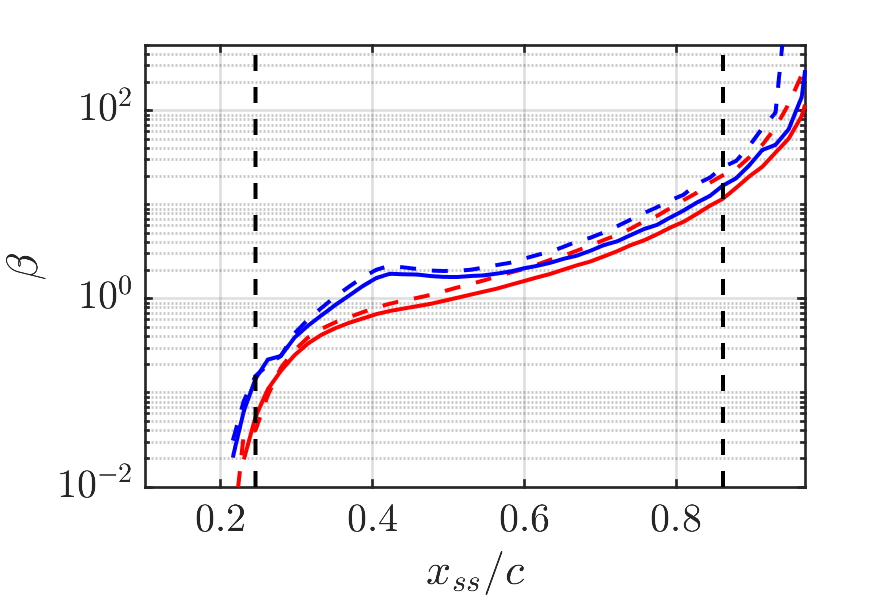}
\end{center}
 \caption{Pressure-gradient parameter $\beta$ on the suction side. Solid and dashed lines are employed for the uncontrolled and the controlled cases, respectively, and blue and red for $Re_c=100,000$ and  $Re_c=200,000$. The vertical dashed lines delimit the control region.}
 \label{fig:beta}
\end{figure}

\begin{figure}[htp]
\begin{center}
   \includegraphics[width=0.22\textwidth]{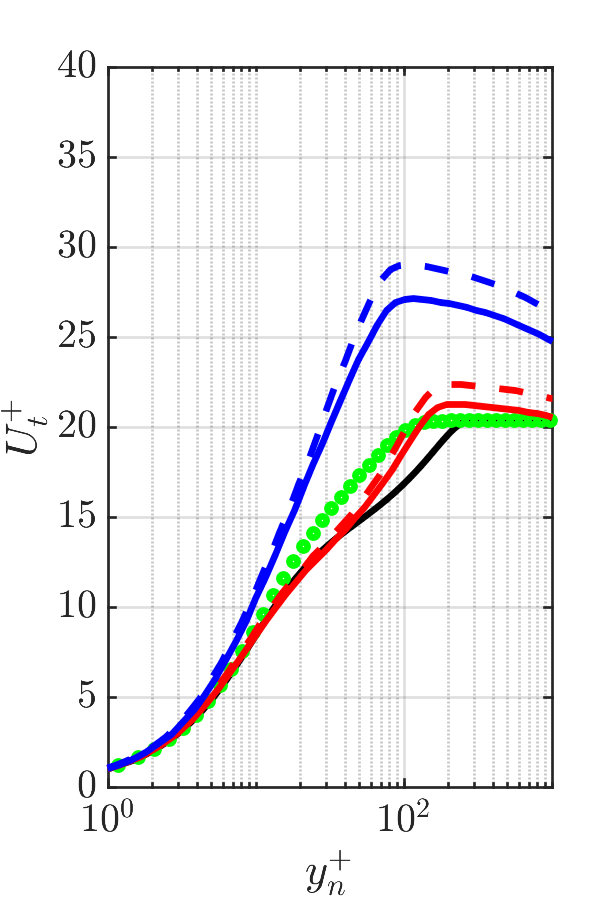}
   \includegraphics[width=0.22\textwidth]{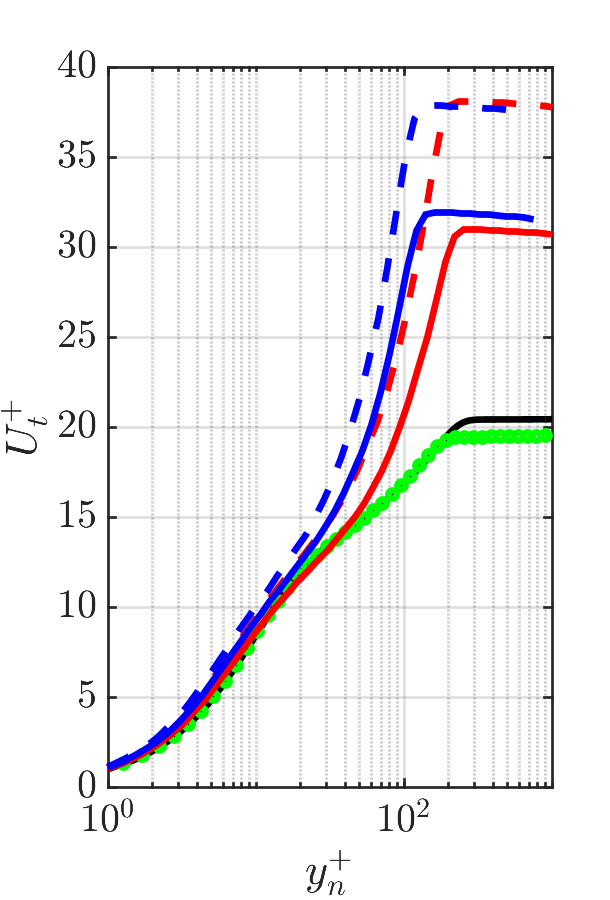}
\end{center}
 \caption{Inner-scaled mean tangential velocity and at (left) $x_{ss}/c=0.4$ and (right) $x_{ss}/c=0.8$. Blue and red are employed for the suction side of the $Re_c=100,000$ and $Re_c=200,000$ cases, respectively, and solid and dashed lines denote the uncontrolled and the controlled cases. The black lines are reference data of ZPG TBL at comparable friction Reynolds number (Schlatter and \"Orl\"u, 2010) and green circles are employed for the pressure side at $Re_c=200,000$.}
 \label{fig:velocity}
\end{figure}

\begin{figure}[htp]
\begin{center}
  \includegraphics[width=0.35\textwidth]{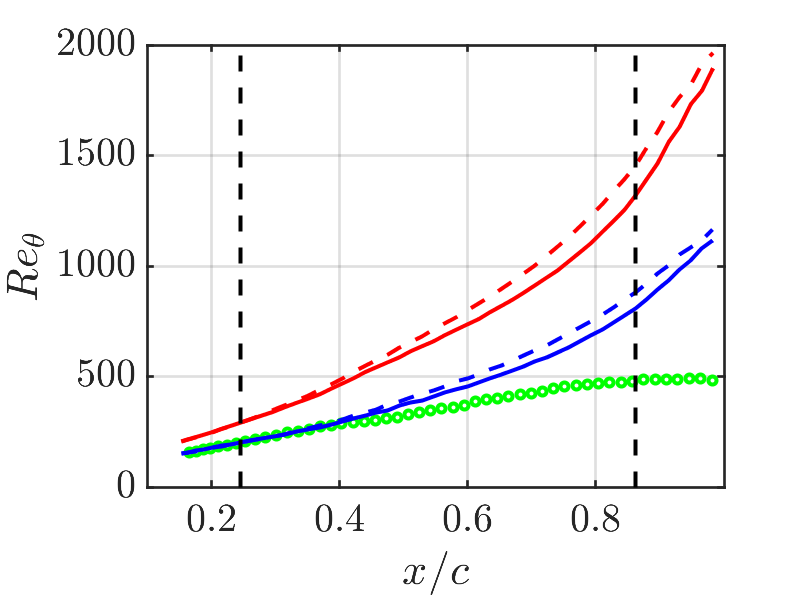}
  \includegraphics[width=0.35\textwidth]{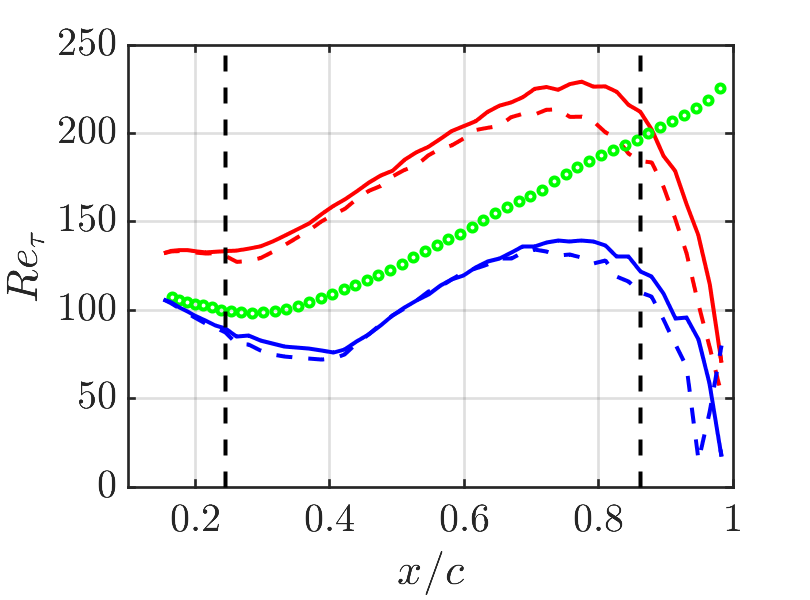}
\end{center}
 \caption{Streamwise evolution of (top) Reynolds number based on the momentum thickness and (bottom) based on the friction velocity. Colors and symbols as in Fig.~\ref{fig:velocity}.}
 \label{fig:reth}
\end{figure}

\begin{figure}[htp]
\begin{center}
 \includegraphics[width=0.35\textwidth]{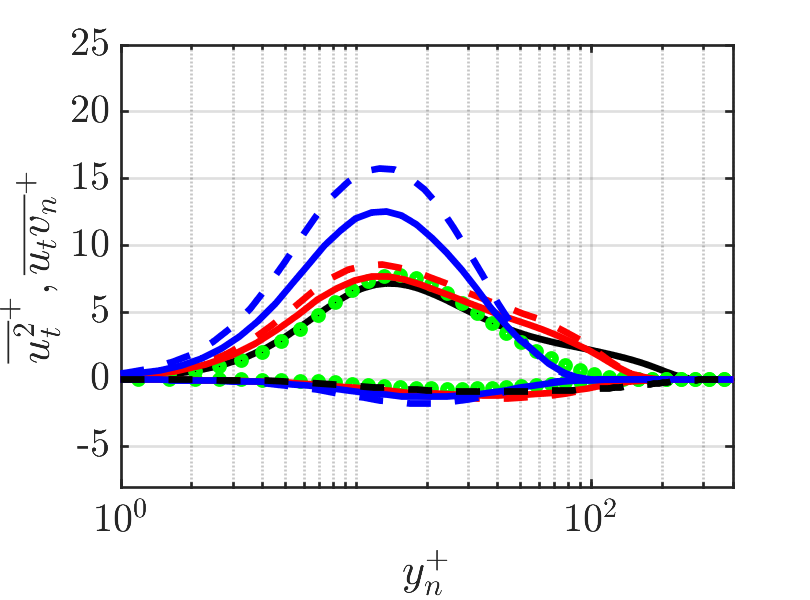}
 \includegraphics[width=0.35\textwidth]{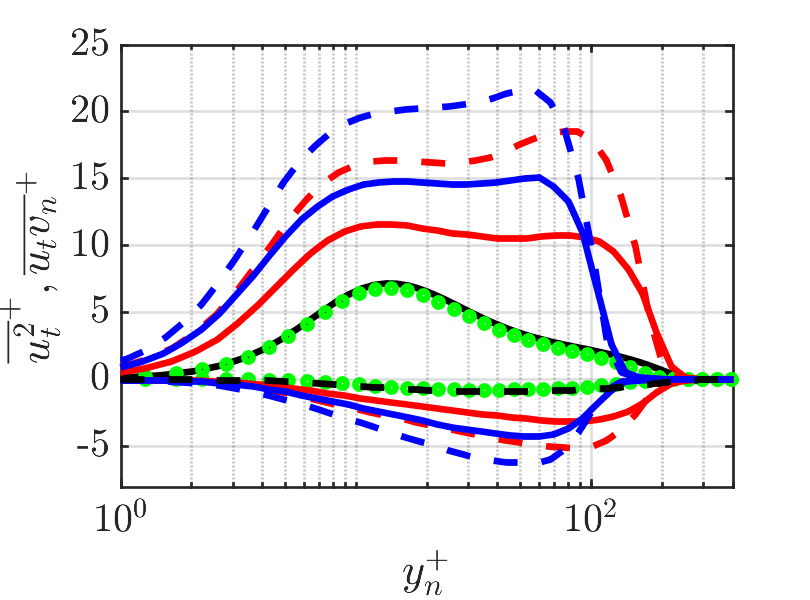}
\end{center}
 \caption{Selected components of the inner-scaled Reynolds-stress tensor at (top) $x_{ss}/c=0.4$ and (bottom) $x_{ss}/c=0.8$. Colors and symbols as in Fig.~\ref{fig:velocity}.}
 \label{fig:Re_stress}
\end{figure}

\begin{figure}[htp]
\begin{center}
  \includegraphics[width=0.22\textwidth]{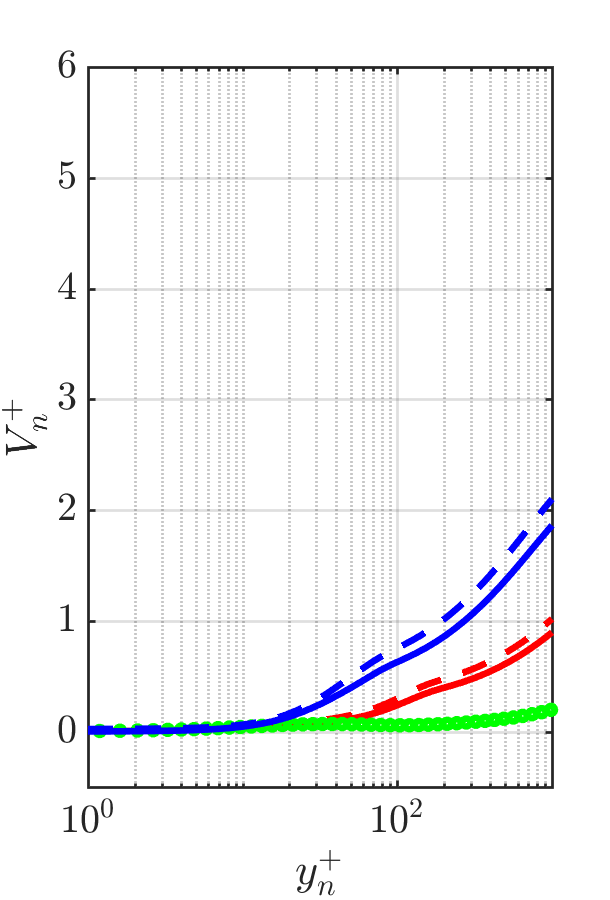}
  \includegraphics[width=0.22\textwidth]{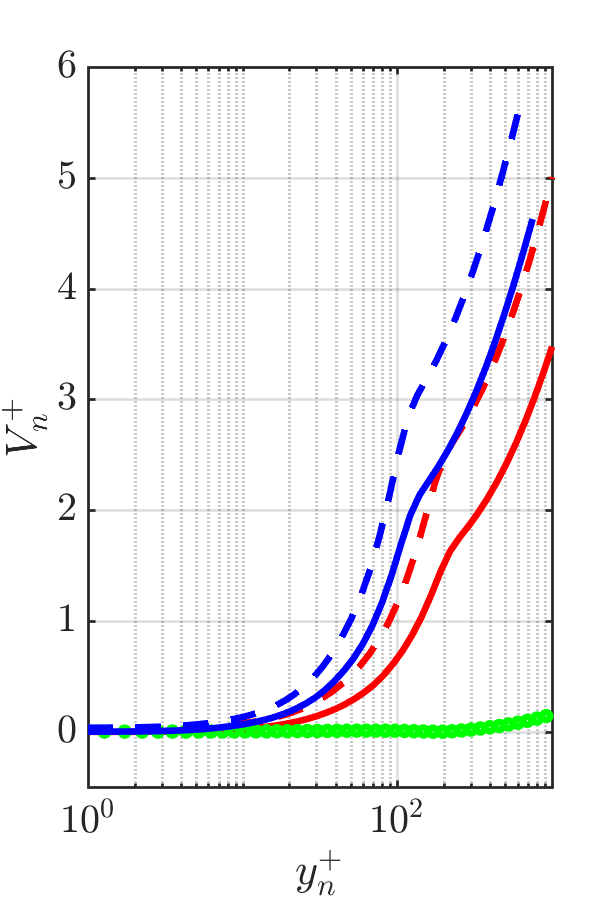}
  \includegraphics[width=0.35\textwidth]{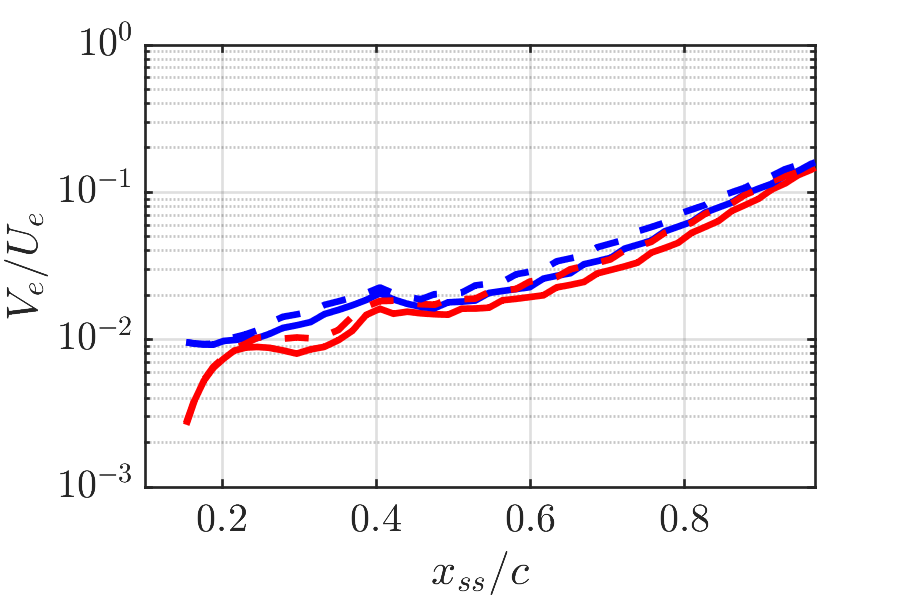}
\end{center}
 \caption{Inner-scaled mean wall-normal velocity velocity at (top-left) $x_{ss}/c=0.4$ and (top-right) $x_{ss}/c=0.8$ and (bottom) streamwise evolution of the outer-scaled wall-normal edge velocity. Colors and symbols as in Fig.~\ref{fig:velocity}.}
 \label{fig:Vvelocity}
\end{figure}

\begin{figure}[htp]
\begin{center}
 \includegraphics[width=0.35\textwidth]{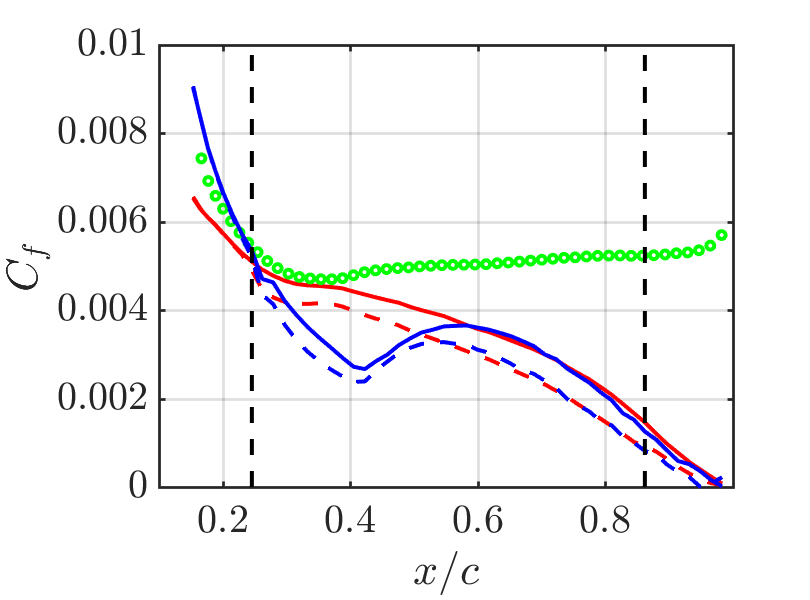}
 \includegraphics[width=0.35\textwidth]{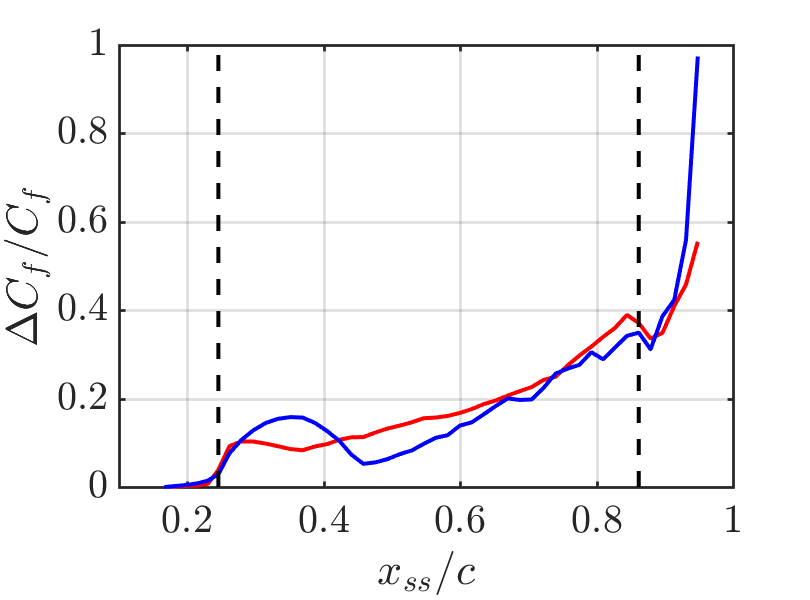}
\end{center}
 \caption{(Top) Streamwise evolution of the skin-friction coefficient $C_f$ and (bottom) relative difference between the $C_f$ from the uncontrolled and the controlled cases. Colors and symbols as in Fig.~\ref{fig:velocity}.}
  \label{fig:Cf}
\end{figure}

\begin{figure}[ht]
\begin{center}
 \includegraphics[width=0.35\textwidth]{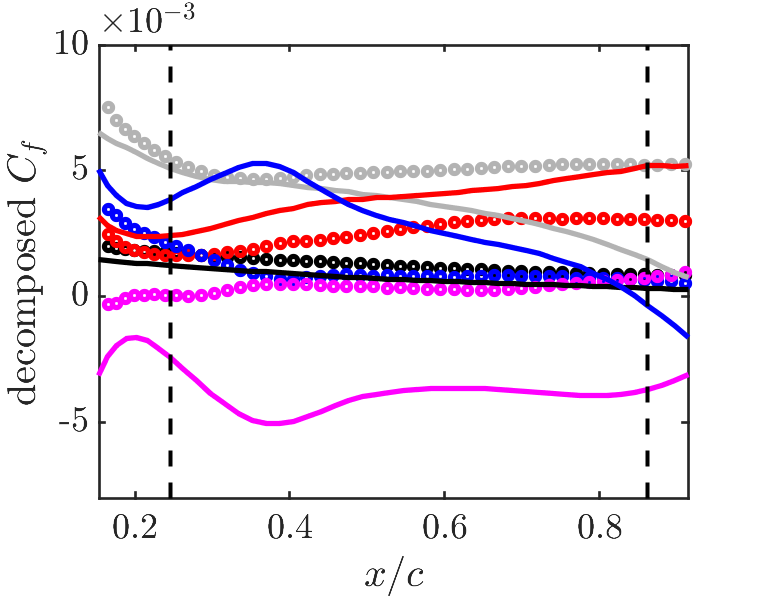}
 \includegraphics[width=0.35\textwidth]{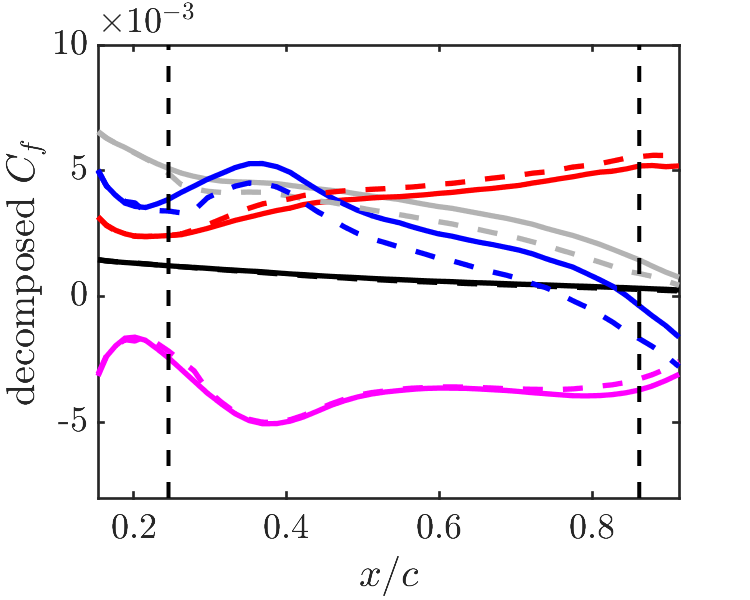}
\end{center}
\caption{FIK decomposition of the skin-friction coefficient for the $Re_c=200,000$ case. (Top) Solid lines and circles denote suction and pressure sides  in the uncontrolled case. (Bottom) Solid and dashed lines denote uncontrolled and controlled cases, both on the suction side. The total $C_f$ is in grey, and the different terms are: $C^\delta$ boundary-layers thickness (black), $C^T$ Reynold-shear stress (red), $C^D$ streamwise development (blue) and $C^P$ pressure gradient (magenta).}
 \label{fig:fik}
\end{figure}

In the present work we assess the impact of uniform blowing on the adverse-pressure-gradient (APG) turbulent boundary layer (TBL) developing on the wing suction side. The NACA4412 profile has been chosen because the suction side is subjected to an APG distribution with moderate dependency on the Reynolds number (Pinkerton, 1938 and Vinuese \emph{et al.}, 2018). The Clauser pressure-gradient parameter $\beta$ represents the ratio between the pressure gradient and the friction forces (Clauser, 1954) and it is defined as $\beta=\delta^*/\tau_w {\rm d}P_e/{\rm d}x_t$, where $\delta^*$ is the displacement thickness and ${\rm d}P_e/{\rm d}x$ is the derivative of the pressure in the tangential direction. The diagnostic scaling is employed to identify the boundary-layer edge, as in Vinuesa \emph{et al.} (2016). The resulting $\beta$ curves are shown in Fig.~\ref{fig:beta} for the suction side for both Reynolds numbers under study.
It can be observed that for the controlled cases $\beta$ is higher, which is a consequence of both $\delta^*$ being larger and $\tau_w$ smaller. 
For both cases, the blowing has an amplitude of $0.1\%$ of $U_\infty$ and it is localised approximately in $0.25<x_{ss}/c<0.86$ (where $ss$ denotes suction side). The controlled region is chosen to be as long as possible to study how blowing affects the development of the boundary layer, but it has been limited upstream to avoid the region directly influenced by the tripping, and downstream to reduce the chance of inducing separation. In Fig.~\ref{fig:velocity} we show the inner-scaled mean velocity profile at $x_{ss}/c=0.4$ and $x_{ss}/c= 0.8$ before and after applying the control. The solid black lines are reference data of zero-pressure-gradient (ZPG) turbulent boundary layer (TBL) at similar friction Reynolds number, which are part of the direct numerical simulation data-set presented by Schlatter and \"Orl\"u (2010).
As discussed by Vinuesa \emph{et al.} (2018), the case at $Re_c=100,000$ is dominated by low-Reynolds-number effects, a fact that explains the peculiar behaviour of the velocity profile at $x_{ss}/c\simeq0.4$. At $x_{ss}/c\simeq0.8$, because of the strong adverse pressure gradient ($\beta\simeq8$), the inner-scaled velocity in the wake region is higher and the profiles do not show a clearly identifiable logarithmic region, at least at these low Reynolds numbers.
The $Re_\theta$ and $Re_\tau$ curves, where $Re_\theta$ is the Reynolds number based on the momentum thickness and $Re_\tau$ the Reynolds number based on the friction velocity, are shown in Fig.~\ref{fig:reth}. 
In Fig.~\ref{fig:Re_stress} we show selected component of the Reynolds-stress tensor at the same location as in Fig.~\ref{fig:velocity}. This figure shows, on the one hand, the increased tangential velocity fluctuations and Reynolds-shear stresses due to the APG; in particular, the tangential fluctuations exhibit an outer peak at $y^+_n \simeq 100$. \\
An additional feature of APG TBLs is the non-negligible values of the mean wall-normal component of the velocity, $V_n$. As reported by Vinuesa \emph{et al.} (2018) for the cases under study, $V_n$ exhibits a dependency on the Reynolds number when scaled in inner-units, decreasing as the Reynolds number increases (as shown in Fig.~\ref{fig:Vvelocity}, top). 
At the same time, $V^+_n$ increases with the streamwise coordinate, a fact that is due to the increase of $\beta$. Considering the controlled cases, it appears that the effect of blowing is stronger approaching the trailing edge, a fact which is in accordance with what was previously observed for the inner-scaled mean tangential velocity. However, the relative effect of the blowing on the mean wall-normal velocity is not uniform in $x$, as it can be illustrated by considering the evolution of the outer-scaled wall-normal edge velocity in Fig.~\ref{fig:Vvelocity} (bottom), defined as $V_e/U_e$ (where $U_e$ the local edge velocity). The discrepancy between the controlled and the uncontrolled cases is approximately uniform up to $x_{ss}/c\simeq0.8$, the point after which it decreases. This suggests that the effect of blowing on the TBL is qualitatively different for different values of the pressure-gradient parameter $\beta$, at least for the Reynolds number under consideration. \\
Summarising, in the controlled cases, the wall-normal convection is increased, the wall-shear stress is reduced and the Reynolds-stress components increase, in particular in the outer region. Consequently, the inner-scaled edge velocity increases, as well as the momentum and displacement thicknesses, a fact that leads to higher $Re_\theta$ and $Re_{\delta^*}$. At the same time, the Reynolds number based on the friction velocity $Re_\tau$ is reduced, together with the skin friction. From this point of view, it appears that the effect of uniform blowing on the flow is similar to that of a strong APG.\\
The impact of the control on the skin-friction coefficient $C_f$, defined as $C_f= 2\tau_w/(\rho U^2_e)$, is shown in Fig.~\ref{fig:Cf} (top). 
The non-monotonic behaviour of $C_f$ for $Re_c=100,000$ is again a low-Reynolds-number effect. The skin-friction reduction is not uniform in $x$, but it increases approaching the trailing edge (where the APG is stronger), and it is observed even downstream of the controlled region. In the $Re_c=100,000$ case, the control also produced a very small separated region in the proximity of the trailing edge. \\
In Fig.~\ref{fig:Cf} (bottom) we report the relative reduction due to the control.
The skin friction can be studied by decomposing it as described by Fukagata \emph{et al.} (2002), in the so-called FIK decomposition:
 \begin{equation}
     C_f(x) = C^\delta(x) + C^T(x) + C^D(x) + C^P(x) \,,
 \end{equation}
where: $C^\delta = 4(1-\delta_{99}/\delta^*)/Re_\delta$ takes into account the boundary-layer thickness ($Re_\delta=U_e \delta_{99} / \nu$); $C^T_f=2\int_0^1 2(1-\eta)(-\overline{u_t v_n}) {\rm d}\eta$ is the contribution from the Reynolds-shear stress; $C^D_f=-2\int^1_0(1-\eta)^2I_x {\rm d}\eta$ is related to the inhomogeneity in the streamwise direction with $I_x=\partial_{x_t}(U_t \,U_t) + \partial_{x_t} \overline{u_t u_t} +\partial_\eta(U_t\, V_n) - (1/Re_\delta)\partial^2_{x_t} U_t$; and $C^P=-2\int_0^1 (1-\eta)^2 \partial_{x_t} P$ (see also Kametani \emph{et al.}, 2015). In the present case, the vertical coordinate in the original definition is substituted by the outer-scaled wall-normal coordinate $\eta=y_n / \delta_{99}$. In Fig.~\ref{fig:fik} we show the result of the FIK decomposition for the $Re_c=200,000$ case on the pressure and suction sides of the uncontrolled wing (top) and on the suction side before and after the control (bottom).  By comparing the pressure and suction sides, it is possible to isolate the effects of the adverse pressure gradient. The boundary-layer term slightly decreases on both sides of the wing, due to the increase in boundary-layer thickness with $x$. The Reynolds-shear stress term increases on the suction and pressure sides with $x$ due to the progressively stronger fluctuations associated to the higher local Reynolds number observed as one moves downstream. The development term presents a more complex behaviour, since it is higher on the suction than on the pressure side up to $x/c\simeq0.8$, the point after which it decreases on the top surface and it becomes eventually negative close to the trailing edge, due to high wall-normal velocity associated to large values of $\beta$. Finally, the pressure term significantly decreases with $x$ on the suction side, due to the fact that the wall-normal velocity and the streamwise pressure gradient are not negligible. \\
The blowing effect is rather different than that of the APG from this perspective. As for APG, the boundary-layer term does not change appreciably and the Reynolds-shear stress term increases because of the energising effect of the blowing on the fluctuations in the outer region. However, because of the wall-normal component of the velocity, the development term decreases while the pressure term does not change appreciably. The net effect is a reduction in $C_f$, since the decrease due to the development term overcomes the increase of the skin friction due to turbulence. \\
Despite the fact that the results above are interesting from a fundamental perspective, in order to understand the interaction between blowing and APG with non-uniform $\beta$ distribution, from an industrial point of view it is necessary to consider how the control strategy modifies the aerodynamic properties of the wing profile. To this end, we compute the lift and drag coefficients as $C_l = 2f_l / (\rho U^2_\infty c )$ and $C_d =2 f_d / (\rho U^2_\infty c)$, where $f_l$ and $f_d$ are the lift and drag forces per (spanwise) length unit, respectively. 
\begin{table}[hbt]
    \centering
    \begin{tabular}{cccc}
        \midrule
        $Re_c$ & $\Delta C_f$ cont. reg. & $\Delta C_f$ s.s. & $C_l/C_d$ \\
        \midrule
        $100,000$ & $16\%$ & $13\%$ & $32\,(34)$  \\ 
        $200,000$ & $17\%$ & $14\%$ & $42\,(44)$  \\
        \midrule
    \end{tabular}
    \caption{Skin-friction reduction over the controlled region and the whole suction side (s.s.) and aerodynamic efficiency for the considered cases (the values in brackets are from the uncontrolled cases).}
    \label{tab:summary}
\end{table}
Table~\ref{tab:summary} reports the integrated skin-friction reduction and the aerodynamic efficiency $C_l/C_d$ for the various cases under consideration. The integrated skin-friction reduction is computed both for the controlled region ($0.24<x/c<0.86$) and for the complete turbulent boundary layer, only excluding the region where separation occurs for the $Re_c=100,000$ controlled case, \emph{i.e.} beyond $x_{ss}/c\simeq 0.95$.
Our results indicate that, despite the reduction in the skin-friction coefficient, the present control not only reduces slightly $C_l$, but also does not decrease $C_d$. The latter is due to the different pressure distribution at the wall induced by the control, which gives to a higher pressure-drag contribution. This effects cannot be observed in studies on simple geometry without curvature effects, such as TBL developing on flat plate or channel flows, where the pressure-drag is zero. This leads to a reduction of the aerodynamics efficiency of $2\%$ for both Reynolds number under study, a fact that highlights the need of exploring a wider range of control configurations, with the aim of maximising the aerodynamic efficiency.
\section{Conclusion and outlook}
In the present work we have discussed an initial assessment of the effects of uniform blowing on the APG TBL over the suction side of NACA4412 airfoil at $Re_c=200,000$, which is characterised by a non-uniform distribution of the Clauser pressure-gradient parameter $\beta$ approximately independent of Reynolds number. Our results show that the effects of blowing on the inner-scaled profiles of the tangential mean velocity and the fluctuations are similar to those of an APG. However, the FIK decomposition shows that the mechanism which leads to a reduced skin-friction is different, since the main contribution to the reduction for APG comes from the pressure-gradient term, while for the blowing control it comes from development term, in particular from the wall-normal velocity. This is in agreement with the previous observations in ZPG TBLs with uniform blowing by Kametani \emph{et al.} (2015). It turns out that the effect of blowing is stronger in the region of the flow subjected to a stronger APG. At the present moment it is not possible to discern whether this is due to the already high wall-normal convection at strong $\beta$, or to the accumulated effects from the upstream region (see for instance Bobke \emph{et al.} (2017) and Vinuesa \emph{et al.} (2017) for additional details regarding the effect of flow history). The present control strategy does not improve the aerodynamic properties of the wing profile. For $Re_c=200,000$, the integrated skin-friction reduction over the controlled region is of $17\%$. However, since the drag coefficient $C_d$ remains almost unchanged and the lift coefficient $C_l$ decreases, the aerodynamic efficiency decreases by $2\%$. \\
Future work will include a comparison between uniform blowing and the body-force damping employed by Stroh \emph{et al.} (2016). The body-force damping will be calibrated to have the same local drag reduction as the one obtained with blowing. Although the two control techniques led to a similar local reduction of the skin friction for ZPG TBLs, they are based on principles which are fundamentally different. The control based on uniform blowing increased the boundary-layer growth, resulting in drag reduction for the uncontrolled region as well, while the body-force damping acts in the opposite way, and the drag is increased in the uncontrolled region. An additional natural extension of this study is to perform a complete parametric analysis aimed at optimising the wing aerodynamic efficiency.

\Acknowledgments
The simulations were performed on resources provided by the Swedish National Infrastructure for Computing (SNIC) at the Center for Parallel Computers (PDC), in KTH, Stockholm. MA acknowledges funding from the Swedish Foundation for Strategic Research, project “In-Situ Big Data Analysis for Flow and Climate Simulations” (ref. number BD15-0082). RV and PS acknowledge the funding provided by the Swedish Research Council (VR) and from the Knut and Alice Wallenberg Foundation. The collaboration between the two research groups was initiated through a Feodor Lynen Research Fellowship of the Alexander von Humboldt for BF. All support is greatly acknowledged. 

%
\begin{References}
\item Bobke, A., Vinuesa, R., \"Orl\"u, R. and Schlatter, P. (2017), History effects and near equilibrium in adverse-pressure-gradient turbulent boundary layers, \emph{J. Fluid Mech.}, Vol. 820, pp. 667--692.
\item Clauser, F. (1954), Turbulent boundary layers in adverse pressure gradients. \emph{J. Aero. Sci.}
Vol. 21, pp. 91--108.
\item Choi, H., Jeon, W.-P. and Kim., J. (2008), Control of Flow Over a Bluff Body. \emph{Annu. Rev. Fluid Mech.}. Vol. 40, pp. 113--139.
\item Fischer, P.F., Lottes, J.W. and Kerkemeier, S.G. (2008), NEK5000: Open source spectral element CFD solver. Available at: \url{http://nek5000.mcs.anl.gov}
\item Fukagata, K., Iwamoto, K. and Kasagi. N. (2002), Contribution of Reynolds stress distribution to the skin friction in wall-bounded flows. \emph{Phys. Fluids}, Vol. 14, pp. 73--76.
\item Gad-el-Hak, M. (2000), Flow Control: Passive, Active, and Reactive Flow Management. \emph{Cambridge University Press}.
\item Jeong, J. and Hussain, F. (1995), On the identification of a vortex. \textit{J. Fluid Mech.} Vol. 285, pp. 69--94.
\item Kametani, Y., Fukagata, K., \"Orl\"u, R. and Schlatter, P. (2015). Effect of uniform blowing/suction in a turbulent boundary layer at moderate Reynolds number. \emph{Int. J. Heat Fluid Flow}, Vol. 55, pp. 132--142.
\item Pinkerton, R.M., (1938). The variation with Reynolds number of pressure distribution over an airfoil section. \textit{NACA Ann. Rep.}, Vol. 24, pp. 65--84.
\item Schlatter, P. and \"Orl\"u, R., (2010). Assessment of direct numerical simulation data of turbulent
boundary layers. \emph{J. Fluid Mech.}, Vol. 659, pp. 116--126.
\item Schlatter, P., and \"Orl\"u, R. (2012), Turbulent boundary layers at moderate Reynolds numbers. Inflow length and
tripping effects. \emph{J. Fluid Mech.} Vol. 710, pp. 5--34.
\item Schlatter, P., Stolz, S. and Kleiser., L. (2004), LES of transitional flows using the approximate deconvolution model. \emph{Int. J. Heat Fluid Flow} Vol. 25, pp. 549--558.
\item Stroh, A., Hasegawa, Y., Schlatter P. and Frohnapfel, B. (2016), Global effect of local skin friction drag reduction in spatially developing turbulent boundary layer, \emph{J. Fluid Mech.}, Vol. 805, pp. 303--321.
\item Vinuesa, R., Bobke, A., \"Orl\"u, R. and Schlatter, P. (2016), On determining characteristic length
scales in pressure-gradient turbulent boundary layers. \emph{Phys. Fluids}, Vol. 28 p. 055101.
\item Vinuesa, R., Hosseni, S.M., Hanifi, A., Henningson, D.S., and Schlatter P. (2017). Pressure-gradient turbulent boundary layers developing around a wing section. \emph{Flow Turbul. Combust.}, Vol. 99, pp. 613--641.
\item Vinuesa, R., Negi, P.S., Atzori, M., Hanifi, A., Henningson, D.S., and Schlatter, P. (2018), Turbulent boundary layers around wing sections up to $Re_c=1,000,000$. \emph{Int. J. Heat Fluid Flow}, Vol. 72, pp. 86--99.
\item Vinuesa, R., \"Orl\"u, R., Vila, C.S., Ianiro, A., Discetti, S., and Schlatter, P. (2017),  Revisiting history effects in adverse-pressure-gradient turbulent boundary layers, \emph{Flow Turbulence Combust}, Vol. 99, pp. 565-587.
\item Vinuesa, R. and Schlatter, P. (2017). Skin-friction control of the flow around a wing section through uniform blowing. \emph{ETC16, 21-24 August, Stockholm, Sweden}.

\end{References}
\end{document}